\documentclass[twocolumn,prl,amssymb,epsfig,aps]{revtex4}

\usepackage{epsfig}

\begin{document}
\textheight 240mm

\title{Reduced carbon solubility in Fe nano-clusters and implications for the growth of single-walled carbon nanotubes}

\author{A. R. Harutyunyan$^{1,*}$, N. Awasthi$^2$, A. Jiang$^2$, W. Setyawan$^2$, E. Mora$^1$, T. Tokune$^1$, K. Bolton$^3$, S. Curtarolo$^{2,*}$} 
\affiliation{
$^1${\footnotesize Honda Research Institute USA Inc. 1381 Kinnear Road  Columbus OH 43212}\\
$^2${\footnotesize Department of Mechanical Engineering and Materials Science Duke University Durham NC 27708}\\
$^3${\footnotesize University College of Bor\aa s SE-501 90 Bor\aa s and Physics Department G\"oteborg University SE-412 96 G\"oteborg Sweden}\\
$^*${\footnotesize corresponding authors: aharutyunyan@oh.hra.com, stefano@duke.edu}
}

\date{\today}

\begin{abstract}
Various diameters of alumina-supported Fe catalysts are used to grow single-walled 
carbon nanotubes (SWCNTs) with chemical vapor decomposition. 
We find that the reduction of the catalyst size 
requires an increase of the minimum temperature necessary for the growth.
We address this phenomenon in terms of solubility of C in Fe nanoclusters
and, by using first principles calculations, we devise a simple model to predict the 
behavior of the phases competing for stability in Fe-C nano-clusters at low temperature.
We show that, as a function particles size, there are three scenarios
compatible with steady state-, limited- and no-growth of SWCNTs, 
corresponding to unaffected, reduced and no solubility of C in the particles. 
The result raises previously unknown concerns about the growth feasibility of 
small and very-long SWCNTs within the current Fe CVD technology, and suggests new
strategies in the search of better catalysts.

\end{abstract}
\pacs{61.46.Df, 65.80.+n, 64.70.Dv, 82.60.Qr}
\maketitle

Among the established methods for single-walled carbon nanotubes (SWCNTs) synthesis \cite{exp1,exp2},
the low temperature catalytic chemical vapor decomposition (CCVD) technique 
is more appropriate for growing nanotubes on a substrate at a target position and, thereby, can
accelerate the integration of this unique material in hybrid electronics.
Reported synthesis of SWCNTs by CCVD at reactor temperature as low as 
$\sim$350-450$^\circ$C was achieved by using hydrocarbons 
with extremely exothermic catalytic decomposition reaction \cite{exp3,exp4}, 
which may release significant free energy (e.g. for acetylene 260 kJ/mol at 650$^\circ$C \cite{exp5}), 
affecting the temperature of the catalyst under certain conditions. 

Considering the vapor-liquid-solid model (VLS) as the most probable mechanism for SWCNT growth \cite{exp8,exp9,exp9b},
an alternate approach for lowering the growth temperature can be the reduction of the catalyst size, since
the Gibbs-Thompson model predicts a decrease of the melting temperature with decreasing cluster size \cite{exp10,the1,the2} 
and the synthesis temperature is correlated to the catalyst-carbon melting and eutectic points \cite{exp11,exp12,exp13}.
While small catalyst particles nucleate small diameter tubes
(this may lead to the diameter control growth, another fundamental problem that hinders nanotube application), 
they also affect 
the morphology of the formed carbon structures \cite{exp14},
the kinetics of the growth \cite{exp15,exp16},
and the solubility of carbon available for the growth process, 
which requires understanding of the widely debated question about the property 
of the catalyst particles, associated with the fundamental problem of their thermodynamic state \cite{the1}.
The main challenges are the un-exposed peculiar features of small catalysts and 
their binary phase diagrams with carbon. 
In this work, we address the carbon solubility problem in Fe nanoparticles
by studying size-dependent growth of SWCNTs by CCVD, and 
the thermodynamics of competing phases in Fe-C clusters with {\it ab initio} modeling.

In our experiments, Fe catalysts supported on alumina powder were prepared using a
the common impregnation method \cite{exp9b}.
To find the minimum synthesis temperature, T$^{min}_{synth}$, as a function of catalyst size,
different particles' dimensions were obtained by varying the Fe:Al$_2$O$_3$
molar ratio 1:15, 1:25, 1:50 and 1:100, \cite{exp17} corresponding
to particles of diameters
$\sim3\pm0.6, \sim2\pm0.8, \sim1.4\pm0.7$, and $\sim1.3\pm0.7$ nm, respectively.
The growth of SWCNTs was performed with CCVD at temperatures between 650-900$^\circ$C, for 90 min over
the pre-reduced catalysts \cite{exp18}. 
To avoid increasing the catalyst temperature,
methane was used as the carbon source since the formation of SWCNTs from methane is not exothermic.
For each case T$^{min}_{synth}$ was determined by analyzing the Raman spectra (Thermo Nicolet
Almega Raman spectrometer equipped with a CCD detector, laser excitations
of 532 and 785 nm) of the samples collected after synthesis at various temperatures.
\begin{figure}[hb]
 \begin{center}
   \vspace{-3mm}
   \centerline{\epsfig{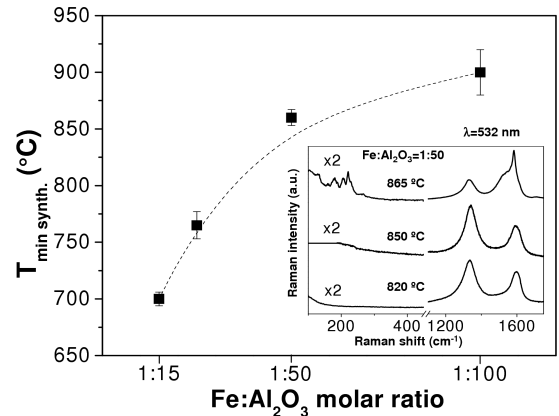}}
   \vspace{-5mm} 
   \caption{\small 
     Evolution of minimum  synthesis  temperature  (T$^{min}_{synth}$) for growing  SWCNTs
     with Fe:Al$_2$O$_3$ molar ratio. The inset shows the Raman spectra for
     samples obtained with a catalyst molar ratio of 1:50 at 820, 850 and 865$^\circ$C.}
   \label{figure1}
 \end{center}
\end{figure}

Figure \ref{figure1} shows the T$^{min}_{synth}$ dependence on the Fe:Al$_2$O$_3$ molar ratio.
The inset presents a typical example of Raman spectra after the synthesis between 820 and 865$^\circ$C.
As can be seen from molar ratio of 1:50, 
SWCNTs can be grown at 865$^\circ$C but not at 850$^\circ$C even though carbon deposition is observed.
In this case, T$^{min}_{synth}$ was estimated to be $\sim$865$\pm$9$^\circ$C.
We find that T$^{min}_{synth}$ increases with decreasing catalyst size 
(molar ratio), contrary to what may be expected from the Gibbs-Thompson model \cite{the1}.
This observation indicates that decomposition of the hydrocarbon alone is not 
enough to grow nanotubes and that the temperature must be increased to ensure that a certain amount 
of carbon dissolves into the particle (considering that the maximum solubility of 
C in Fe depends on the catalyst size, as shown later).
In fact, temperature must be increased to overcome the loss of solubility of C in the 
catalytically-active phase competing for thermodynamical stability with a nucleating carbide,
and not only to enhance diffusion of C (otherwise below T$^{min}_{synth}$  we would have shorter nanotubes, instead of their absence).

We believe that the origin of this apparent paradox lies in a 
novel phenomenon, i.e., a reduced solubility of C in Fe nanoparticles.
Within the VLS framework with bulk diffusion as the rate-limiting step \cite{VLS,Baker72,Baker75,exp9b}, 
this implies an increase of temperature 
to achieve comparable amount of dissolved carbon to allow growth.
In Ref. \cite{the1} we have shown that the eutectic point $(x^{C}_{eut},T_{eut})$ of Fe-C clusters shifts 
toward lower carbon concentrations, $x^{C}_{eut}$, with decreasing particle size (Figure 8 of \cite{the1}).
Due to the high energetic cost for bringing bulk cementite off-stoichiometry
(in the Fe-C phase diagrams Fe$_3$C forms two-phase regions with austenite ($\gamma$) 
and ferrite ($\alpha$) without going off-stoichiometry \cite{pauling}), 
the most probable cause of the shift of $x^{C}_{eut}$ is a reduced solubility of C.
This does not necessarily imply that the total amount of C present in the nanoparticle decreases, 
because solubility counts only the carbon dissolved randomly in the solid Fe-rich phases,
(in equilibrium with Fe$_3$C (below $T_{eut}$) or with the liquid/viscous phase (above $T_{eut}$)).

The accurate analysis of the phenomenon can be achieved by calculating the interplay between
the phases competing for stability at the temperatures of the process. 
For nanoparticles, this task is generally unsolvable, although qualitative 
information can be extracted from approximate zero-temperature first-principles modeling.
In such approaches, by comparing the formation energies of the candidate phases, 
we determine the stability of the system at low-T and give indications for higher temperature behavior.
The {\it ab initio} simulations presented here are performed with {\small VASP} \cite{kresse1993},
using projector augmented waves (PAW) \cite{bloechl994} and exchange-correlation functionals as 
parameterized by Perdew, Burke, and Ernzerhof (PBE) \cite{PBE} for the generalized gradient approximation (GGA). 
Simulations are carried out with spin polarization, at zero temperature, and without zero-point motion.
All structures are fully relaxed.  Numerical convergence to within about 2 meV/atom is ensured by 
enforcing a high energy cut-off (500 eV) and dense {\bf k}-meshes.

\begin{figure}[htb]
 \begin{center}
 \centerline{\epsfig{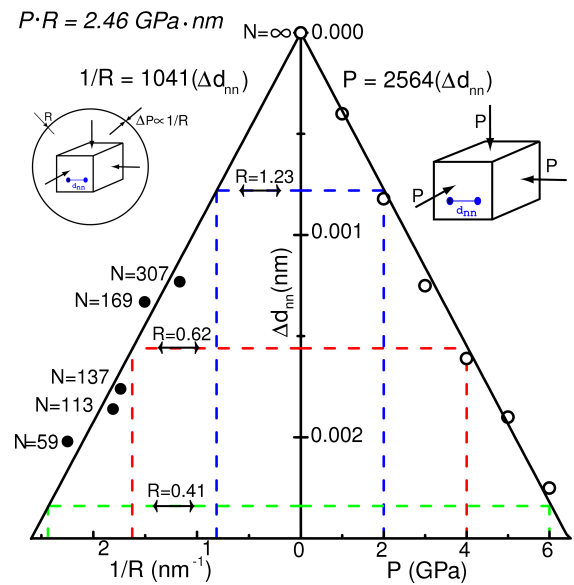}}
 \vspace{-2mm} 
 \caption{\small (color online).
 Size-pressure approximation for Fe nanoparticles. 
 Given a spherical particle, 
 we estimate the hydrostatic pressure due to the surface curvature 
 by calculating the deviation of the average bond length inside the cluster,
 $\Delta d_{nn} \equiv d_{nn}^{0}-d_{nn}$, 
 and comparing it to the deviation of the bond length for the bulk material 
 as a function of hydrostatic pressure. }
 \vspace{-8mm}
\label{figure2}
\end{center}
\end{figure}

Our model is based on the following assumptions: 
(i) {\it Mechanism.}
The behavior of carbon is determined by the interplay of 
four competing phases as a function of catalyst size:
pure bcc-Fe,
C dissolved in ferrite ($\alpha$-FeC$_x$),
ordered cementite (Fe$_3$C), 
and carbon SWNTs \cite{exp9b,pauling}.
The pure-Fe phase is taken to be bcc because our simulations are aimed
to explore the low temperature regime of catalytic growth.
The $\alpha$-FeC$_x$ phase is simulated by taking samples of bcc supercells 
with different concentrations of interstitial carbon
(Fe$_{32}$C, Fe$_{24}$C, Fe$_{16}$C). 
We are not required to generate truly random phases by using 
the special quasirandom structure formalism, because the low concentration of carbon,
guarantees enough distance between replica of C to achieve convergence \cite{SQS1,SQS2}. 
In addition, higher concentrations of C are not required to be explored, 
because even in bulk $\alpha$-FeC$_x$ the solubility is small \cite{pauling}.
(ii) {\it Carbon source.} 
Free carbon atoms come from the dissociation of the feed-stock 
on the surface of pure-Fe and random FeC$_x$ catalysts only.
Formation of cementite stops the process due to its different 
activity and diffusion properties (as show in Fig. 1 of \cite{exp9b} 
and references therein).
(iii) {\it Nanotubes diameter.} 
To minimize the curvature energy of the tube, active catalysts produce 
nanotubes that have a similar diameter as the particle (CVD experiments have shown clear correlation between
the two diameters \cite{ntdia5}).
(iv) {\it Size-pressure approximation.} 
In nanoparticles, surface curvature and superficial dangling bonds
are responsible for internal stress fields which modify the atomic bond lengths inside the particles. 
For spherical clusters, the induced stress fields are hydrostatic and the compressions are isotropic.
The phenomenon can be modeled with the Young-Laplace equation
$\Delta p=2\gamma/R $ where the proportionality constant $\gamma$ (surface tension for liquid particles) 
can be calculated with {\it ab initio} methods.
As a first approximation, by neglecting all the surface effects not included in the curvature, 
the study of phase diagram for spherical particles can be mapped onto the study 
of phase diagrams for bulk systems under the same hydrostatic pressure produced 
by the curvature, as depicted in Fig. \ref{figure2} \cite{note1}.
It is worth mentioning that our reference for carbon
is taken to be the zero pressure nanotube phase, 
different from the other carbon references used for investigating 
Fe-C under pressure \cite{pressure1,pressure2,pressure3}. 

Figure \ref{figure2} shows the implementation of the ``size-pressure approximation'' 
for Fe nanoparticles.
On the left hand side we show the {\it ab initio} calculations 
of the deviation of the average bond length inside the cluster 
$\Delta d_{nn}\equiv d_{nn}^{0}-d_{nn}$ ($d_{nn}^{0}=0.2455$ nm\,\, is our bulk bond length),
for bcc particles of size $N=59, 113, 137, 169, 307$ and $\infty$ (bulk) as a function of the inverse radius ($1/R$).
The particles were created by intersecting a bcc lattice with different size spheres.
The particle radius is defined as $1/R\equiv 1/N_{scp}\sum_i 1/R_i$ where the sum is taken over 
the atoms belonging to the surface convex polytope ($N_{scp}$ vertices) and 
$R_i$ are the distances to the geometric center of the cluster \cite{note2}.
The left straight line is a linear interpolation between $1/R$ and $\Delta d_{nn}$ calculated with 
the constraint of passing through $1/R=0$ and $\Delta d_{nn}=0$ ($N=\infty$, bulk).
The right hand side shows the {\it ab initio} value of
$d_{nn}$ in bulk bcc Fe as a function of hydrostatic pressure, $P$.
The straight line is a linear interpolation between $P$ and $\Delta d_{nn}$ calculated with 
the constraint of passing through $P=0$ and $\Delta d_{nn}=0$ (bulk lattice).
By following the colored dashed paths indicated by the arrows we can map the 
analysis of nanoparticles' stability as function of $R$ onto bulk stability as function of $P$, 
and obtain the relation between the radius of particle/nanotube and the effective pressure $P\cdot R=2.46$ GPa$\cdot$nm.
It is important to mention that our $\gamma=1.23$ J/m$^2$ is not a real surface 
tension but an {\it ab initio} fitting parameter describing size-induced stress in nanoparticles.
In addition, it compares well with the experimental value of the surface tension 
of bulk Fe at its melting point $\sim1.85$ J/m$^2$
\cite{gammaexp}.

\begin{figure}[ht]
 \begin{center}
   \centerline{\epsfig{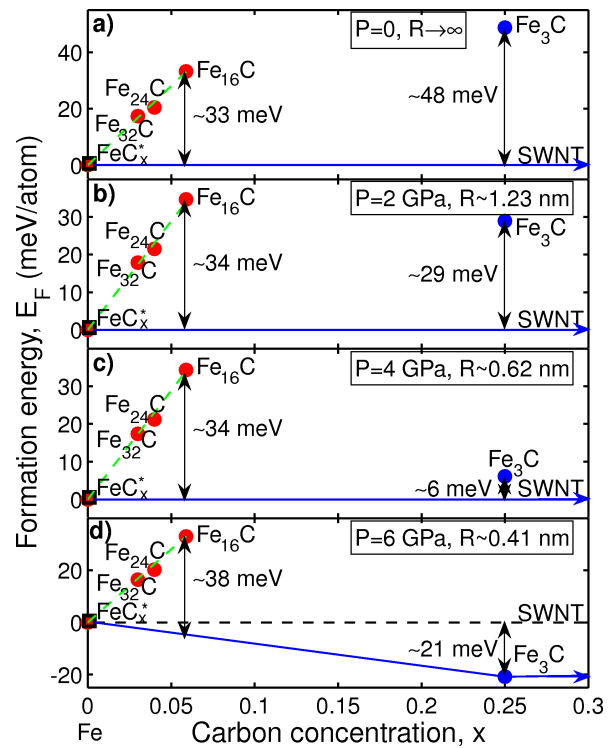}}
 \vspace{-1mm} 
 \caption{\small (color online).
   The scenarios described in the text with their implication in CVD SWCNT growth with Fe nano-catalysts.
   $P$ is the pressure at which the calculations are performed, and $R$ is the radius of the corresponding
   SWCNT within the size-pressure approximation. 
   E$_f$[Fe$_3$C]=E$_f$[$\alpha$-FeC$_{\sim 0.00102}^{sol}$] for $R_{min}\sim 0.58$ nm ($P\sim4.3$GPa).
 }
 \vspace{-12mm}
\label{figure3}
\end{center}
\end{figure}

Figure \ref{figure3} shows the evolution of competing phases as function 
of concentration and particle's size. 
In panels (a) through (d), we plot formation energies,
$E_f[{\rm Fe}_{1-y}{\rm C}_{y}] \equiv E[{\rm Fe}_{1-y}{\rm C}_{y}]-(1-y)E[{\rm Fe}]-yE[{\rm C}]$, 
calculated with respect to the pure constituents of the reaction, bcc-Fe and carbon-SWCNTs 
(with the same diameter as the particle). 
The dashed green lines interpolating Fe$_{16}$C through Fe, are used to estimate 
the formation energies of the random phase with maximum solubility 
of carbon ($\alpha$-FeC$_{\sim 0.00102}^{sol}$, 
black squares near the origin in Fig. \ref{figure3} labeled as FeC$_x^\star$)
corresponding to 0.022 weight\% \cite{pauling}.
A structure at a given composition is considered stable (at zero temperature 
and without zero-point motion) if it has the lowest formation energy for any 
structure at this composition,and, if on the binary phase diagram, it lies below 
the {\it convex hull of tie lines} connecting all the other stable structures \cite{SC13,SC20}.
Phases lying above the convex hull and with small positive formation energies,
might be explored by the thermodynamics of the system through 
configurational and vibrational entropic promotion.

By varying the radius of the particle, the stability of the 
competing phases, $\alpha$-FeC$_x$ and Fe$_3$C, changes considerably.
There are three possible scenarios.
{\it Scenario I} is shown in panels (a) and (b). 
For big particles, $R >> R_{min}$, Fe$_3$C has formation energy higher than the max
solubility phase ($\alpha$-FeC$_{\sim 0.00102}^{sol}$).
Therefore the pollution of carbon at low and medium temperature, 
cannot cause the big particle to undergo phase transition by nucleating cementite.
Hence, such particles remain in the catalytically-active random $\alpha$FeC$_x$ state, 
by keeping their concentration of carbon between 0 to $\sim0.102\%$ \cite{the1} 
(solubility is unaffected), and by implying a balance between in- and out- flows of carbon 
which can guarantee the steady state growth of nanotubes. 
Thermodynamically, in this regime, 
SWCNTs, MWCNTs and carbon fibers could be grown indefinitely and the only
limitation is the availability of carbons feed-stock \cite{Wood_PRB_2007}.
Thus, experiments performed with particles of these sizes would
be described by Arrhenius equations governing catalytic activity 
and diffusion properties.
The minimum radius, $R_{min}$, can be estimated by interpolating
the pressure at which the energy of Fe$_3$C and $\alpha$-FeC$_{\sim 0.00102}^{sol}$ are equal,
and mapping such pressure in the size-pressure relation of Fig. \ref{figure2}. 
We obtain $P\sim4.3$GPa and $R_{min}\sim0.58$nm.
{\it Scenario II} is shown in panel (c). 
For particles of size $R\sim R_{min}$, the max solubility phase and Fe$_3$C have similar formation energies. 
The energetically competing Fe$_3$C causes depletion of C 
in $\alpha$-FeC (reduced solubility), nucleation of ordered Fe$_3$C, 
and overall reduction of catalytically active random Fe.
If exposed to hydrocarbons at elevated temperatures, 
such particles would be capable 
of dissociating carbon and growing SWCNTs with concomitant nucleation of cementite.
Such nucleation slowly poisons the particle and terminates the growth. 
In this regime, SWCNTs can be produced up to a certain critical length
depending on the net flow of carbon. 
By varying $R$, the critical length goes from infinity ($R>R_{min}$) to zero ($R< R_{min}$).
Experiments performed with particles of size $R\sim R_{min}$ would show 
on/off growth at low temperature and Arrhenius behavior at high temperature.
{\it Scenario III} is shown in panel (d). By further reducing the size of the particle, $R<< R_{min}$,
the formation energy of cementite becomes negative 
($E_f[\alpha$-FeC$_{\sim 0.00102}^{sol}]$ is always very close to zero).
The stability of Fe$_3$C over the range 0 to 25\% carbon in the phase diagram, 
indicates that the nucleation of Fe$_3$C occurs simultaneously with the carbon pollution. 
The particle transforms into Fe$_3$C as rapidly as the availability of feed-stock allows, 
thus no random phase coexists at low temperature 
(max solubility of C is zero), and no out-flow of carbon occurs.
Particles with $R<R_{min}$ cannot grow SWCNTs, 
and $R_{min}$ can be considered as a lower limit for SWCNTs' size
in low-temperature CVD growth with Fe nano-catalysts.
Experiments performed with such particles would result 
in Fe$_3$C nanoparticles and no appreciable nanotube productions.

We address the extension of the model to higher temperatures in a qualitative framework \cite{curie1}.
Increasing the temperature of the reaction and considering fcc Fe nanoparticles 
has two consequences. 
Fe$_3$C becomes stable at medium temperature \cite{pauling} 
and, by reducing particle size, the stability versus size follows 
similar arguments as the case of the bcc reference. 
Thus $R_{min}\rightarrow \infty$ and a steady state SWCNT growth is, {\it a priori}, unobtainable. 
However, since the maximum solubility of C in austenite $\gamma$-FeC$_x$ is bigger 
than in ferrite $\alpha$-FeC$_x$ 
(due to the fact that interstitial holes in the fcc lattice are larger 
than those in the bcc lattices ($\sim$2.0 wt\% in $\gamma$-FeC$_x$ \cite{pauling})),
the life of the catalytic particle can somehow be longer than the previous scenarios 2 and 3, 
and the grown SWCNTs can be quite long. 
The lack of steady state at high temperature can be addressed 
by alloying the nanoparticle with other metals to reduce cementite stability 
and 
by simultaneously promoting other more stable alloyed random phases with considerable catalytic activity.

In this manuscript we have addressed the thermodynamic limits 
of bcc Fe nano-catalysis technology for SWCNT growth with low temperature CVD. 
Using a ``size-pressure approximation'' and zero temperature {\it ab initio} modeling 
we identify ranges of nanoparticle sizes which are compatible 
for steady state-, limited- and no-growth of SWCNTs corresponding to unaffected, 
reduced and no solubility of C in the $\alpha$-FeC$_x$ nanoparticles. 

We wish to acknowledge helpful discussions with 
A. Kolmogorov, H. Baranger, T. Tan, N. Li, A. Ferrari, 
S. Hofmann, F. Cervantes-Sodi, and G. Csanyi. 
The authors are grateful for computer time allocated 
at the Swedish National Supercomputing and at the Teragrid facilities. 
This research was supported by Honda Research Institute USA, Inc. 
SC is supported by ONR (N00014-07-1-0878) and NSF (DMR-0639822).


\begin{thebibliography}{99}
\bibitem{note1}
 Since the amount of C present in the Fe particle is limited, we
 take $\gamma$ to be independent of the C concentration.
\bibitem{note2}
  The radius is so defined because ``$1/R_i$'' are the factors producing internal stress the particle.
\end{thebibliography}
\end{document}